\documentclass[aps,prb,twocolumn,showpacs,groupedaddress]{revtex4}  
\usepackage{graphicx}  
\usepackage{dcolumn}   
\usepackage{bm}        
\usepackage{amssymb,amsmath,dsfont}   
\usepackage{epsfig}

\renewcommand{\Re}{{\rm Re}\,}
\renewcommand{\Im}{{\rm Im}\,}

\newcommand{\no}{:}
\newcommand{\on}{:}
\newcommand{\dd}{{\rm d}}
\newcommand{\xUnit}{{\mathds 1}}

\newcommand{\eq}{{\rm eq}}
\newcommand{\free}{0}

\newcommand{\vF}{v_F}

\DeclareMathOperator{\Det}{Det}

\begin{document}

\title{Interaction Quench in Nonequilibrium Luttinger Liquids}
\author{St\'ephane~Ngo~Dinh,$^{1}$ Dmitry A. Bagrets,$^{2}$ and Alexander
D.~Mirlin$^{1,3,4}$}

\affiliation{$^{1}$Institut f\"ur Theorie der \!Kondensierten \!Materie and DFG Center for Functional Nanostructures, Karlsruhe Institute of Technology, 76128 Karlsruhe, Germany\\
$^{2}$Institut f\"ur Theoretische Physik, Universit\"at zu K\"oln, Z\"ulpicher Str.~77, 50937 K\"oln, Germany\\
$^{3}$Institut f\"ur Nanotechnologie, Karlsruhe Institute of Technology, 76021 Karlsruhe, Germany\\
$^{4}$Petersburg Nuclear Physics Institute, 188300 St.~Petersburg, Russia}

\date{\today}

\begin{abstract}
We study the relaxation dynamics of a {\it nonequilibrium}
Luttinger liquid after a sudden interaction switch-on (``quench''), focussing 
on a double-step initial momentum distribution function. 
In the framework of the non-equilibrium bosonization, the results are obtained in terms of singular 
Fredholm determinants that are evaluated numerically and whose asymptotics are  found analytically.
While the quasi-particle weights decay exponentially with time after the quench, this
is not a relaxation into a thermal state, in view of the integrability of the
model. The steady-state distribution emerging at infinite times retains two edges 
which support Luttinger-liquid-like power-law singularities smeared by dephasing. 
The obtained critical exponents and the dephasing length are found to depend on the
initial nonequilibrium state.
\end{abstract}

\maketitle 

\section{Introduction}

Quantum physics of interacting one-dimensional (1D) systems represents a
fascinating research area \cite{giamarchi-book}. Most important experimental
realizations include electrons in 1D nanostructures (quantum
Hall and topological insulator edges, carbon nanotubes, semiconductor quantum
wires), quantum spin chains, and cold atoms (bosons as well as fermions) in
optical traps. One of central directions of current research is the  physics of
nonequilibrium phenomena in these structures. 

In the cold atoms context
(see Ref.~\onlinecite{Bloch:2008} for a review, the most frequently considered
nonequilibrium setting is a quantum quench
\cite{Polkovnikov:2011,Greiner:2002,Kinoshita:2006,Hofferberth:2007,Gring:2012}:
one explores a quantum evolution of the system after a sudden change of one of
the parameters. In particular, one can modify the optical lattice potential
confining the atoms. An alternative possibility is to suddenly change  the
interaction strength by employing a strong dependence of interaction on the
magnetic field in the vicinity of Feshbach resonance.
It has been recognized that, upon an interaction
quench, a Luttinger liquid shows an interesting dynamics and eventually evolves
into a non-thermal state characterized by nonequilibrium power-law
correlations
\cite{Cazalilla:2006,Rigol:2007,Calabrese:2006}. Subsequent theoretical works
explored various generalizations of this problem
\cite{Kollath:2007,Gangardt:2008,Cramer:2008,Iucci:2009,Rigol:2009,Uhrig:2009,
Manmana:2009,Sabio:2010,Barmettler:2010,Mitra:2011,Foster:2011,
Perfetto:2011,Dora:2011,Karrasch:2012,Calabrese:2011,Nessi:2013}.

On the other hand, in the context of electronic systems, typical
nonequilibrium setups are of steady-state character (as obtained by applying
bias voltages to some of source electrodes). Recent experiments have addressed
nonequilibrium spectroscopy of carbon nanotubes
\cite{Chen09} and quantum Hall
edge states \cite{Altimiras10} as well as nonequilibrium edge state
interferometry
\cite{MarcusWest09,
Yamauchi09,ZhangMarcus09,Heiblum09,heiblum03,roche07,strunk07,%
schoenenberger}. On the theory side, one of important recent theoretical
advances was a development of the method of nonequilibrium bosonization
\cite{Gutman:2010,Gutman:2011,Protopopov:2012,Protopopov:2013} that permits, in
particular, to treat Luttinger liquids with distribution functions of incoming
electrons that have multiple Fermi edges. It was shown that this leads to a
multiple-branch zero-bias anomaly with exponents and dephasing rates controlled
by the nonequilibrium state of the system. Related results have been obtained
for the problems of quantum Hall edge state spectroscopy
\cite{Kovrizhin:2011,Levkivskiy:2012} and Mach-Zehnder interferometry
\cite{Kovrizhin:2009,Levkivskiy:2009,NgoDinh:2013}.

While two types of nonequilibrium Luttinger liquid setups (a temporal
interaction quench and a steady-state with multiple Fermi edges originating from
applied voltages) are quite different, there is a remarkable similarity between
the results. In both cases, one finds non-trivial power-law exponents that are
essentially different from the equilibrium ones. In this paper, we show that
both kinds of nonequilibrium settings can be treated within a unified
framework of the nonequilibrium bosonization. We employ this formalism to
explore the dynamics after an interaction quench in an initially
nonequilibrium Luttinger liquid. The Fermi-edge exponents and dephasing
are controlled by charge fractionalization at temporal (and possibly spatial)
boundaries of the interaction region and by the input nonequilibrium
distributions.

\section{Equilibrium quench in the Luttinger model}
\label{s2}
The Luttinger model describes the low-energy physics of interacting 1D
fermions. It turns out that low-energy excitations can be fully captured in
terms of bosonic modes. A free 1D fermionic system with
right-\mbox{(left-)}moving
modes $\psi_+$ ($\psi_-$) close to the Fermi points can be mapped onto a free 1D
bosonic system with a linear spectrum
\begin{align*}
	H_0 &= -i v_F \no\int\!\dd x\,\left(\psi_+^\dagger \partial_x \psi_+-\psi_-^\dagger \partial_x \psi_-\right)\on\\
	& = \pi v_F \no \int\!\!\dd x\, \left(\varrho_+^2+\varrho_-^2\right)\on
\end{align*}
with density operators $\varrho_\pm (x) =\no \psi_\pm^\dagger(x)
\psi_\pm(x)\on$ and Fermi velocity $v_F$. The fermionic operators can be
likewise
expressed in terms of bosonic operators
\begin{align*}
	\psi_\eta(x) \sim \frac{\eta}{\sqrt{2\pi a}} e^{i\phi_\eta},\quad \eta=\pm,
\end{align*}
where the phase operators $\phi_\eta$ are related to densities via
$\varrho_\eta(x)=(\eta/2\pi)\partial_x\phi_\eta(x)$. While
counterpropagating density modes are uncoupled for originally free fermions,
they  are mixed by interaction. The interaction Hamiltonian for short-range
interaction is
\begin{multline*}
	H_{\rm int}(t) = \frac{g_4(t)}2 \no\int\!\!\dd x\, \left(\varrho_+^2+\varrho_-^2\right) \on\\ +g_2(t) \no \int\!\!\dd x\, \varrho_+(x) \varrho_-(x) \on.
\end{multline*}
To describe the interaction quench we let the coupling parameters
$g_j(t)=g_j\theta(t)$  be time dependent. In more physical terms we assume
that the switching time is much shorter than all characteristic time scales of
the problem set for example by the inverse voltage $U$, see Sec.~\ref{s3.1}
below. In the presence of interaction the
new bosonic eigenmodes, the ``plasmons'' $\tilde \varrho_\eta$, are obtained by
the Bogoliubov transformation
\begin{align} \label{eqn:Bogoliubov}
	\begin{pmatrix}
		\varrho_+\\ \varrho_-
	\end{pmatrix} =
	\begin{pmatrix}
		c & s\\ s & c
	\end{pmatrix}
	\begin{pmatrix}
		\tilde \varrho_+ \\ \tilde \varrho_-
	\end{pmatrix}, \quad
	c \equiv \frac{1+K}{2\sqrt K}, \ s\equiv \frac{1-K}{2\sqrt K}
\end{align}
with the Luttinger parameter
\begin{align*}
	K = \sqrt{\frac{2\pi v_F+g_4-g_2}{2\pi v_F+g_4+g_2}}.
\end{align*}
The full Hamiltonian after the quench then reads
\begin{align}
	H=H_0+H_{\rm int}(t>0)=\pi u \no \int\!\!\dd x\, \left(\tilde \varrho_+^2+\tilde \varrho_-^2\right)\on
\end{align}
with the plasmon velocity
\begin{align*}
	u = v_F \sqrt{\left(1+g_4/(2\pi v_F)\right)^2+\left(g_2/(2\pi
v_F)\right)^2}.
\end{align*}
In thermal equilibrium the many-body density matrix $\hat \varrho={\cal Z}^{-1}
e^{-H/T}$ is a function of $H$, thus it is straightforwardly expressed as an
exponential of bilinears of bosonic fields.

Ref.~\onlinecite{Cazalilla:2006,Iucci:2009} considered time-evolution after a
sudden interaction switch-on. The initially noninteracting system is prepared in
the thermal equilibrium state $\hat \varrho_0={\cal Z}_0^{-1} e^{-H_0/T}$ which
after the quench no longer represents equilibrium (with respect to the full
Hamiltonian $H$). However, the time-evolution of $\hat \varrho_0$,
$\varrho_\eta$ and $\psi_\eta$ can be deduced by the Bogoliubov transformation
(\ref{eqn:Bogoliubov}). 
Calculations there were performed with a finite interaction range $R_0\sim
v_F/\Lambda$  as short-distance regularization. At long distances $\bar x\gg
R_0$ results are insensitive to the regularization scheme, and
momentum-dependent coupling parameters (associated with finite interaction
range) can be replaced by their zero momentum values, $g_j(q)\approx g_j(q=0)$.
The equal-time correlation function then is
\begin{align} \label{eqn:LLEqQGF}
	G^<_{\eq+}(\bar x,\bar t;0,\bar t)= G^<_{\free+}(\bar x,0) \left\lvert 
\frac{R_0}{\bar x}\right\rvert^{{\tilde \gamma}^2} \left\lvert \frac{(2u\bar
t)^2-\bar x^2}{(2u\bar t)^2}\right\rvert^{{\tilde \gamma}^2/2},
\end{align}
where the exponent is determined by $\tilde\gamma\equiv (1-K^2)/{4K}$ and
$G_{0+}^<(\bar x,0)$ is the free fermionic Green's function.

For short times such that $2u\bar t\ll |\bar x|$ the correlation function
$G^<_+(\bar t;\bar x,0)\approx Z(\bar t) G^<_{0+}(\bar x;0)$ can be interpreted
as the Green's function of an effective time-dependent Fermi liquid with
``Landau quasiparticle weight''
\begin{align} \label{eqn:QuenchEqQPWeight}
Z(\bar t)=\left({R_0}/{2u\bar t}\right)^{{\tilde\gamma}^2}
\end{align}
which gives rise to a discontinuity in the momentum distribution function
$n_+(p)$ at Fermi momentum $p=p_F$. According to (\ref{eqn:QuenchEqQPWeight})
the jump decays algebraically with time~$\bar t$.

For large times $\bar t\to \infty$ the system reaches a time-independent steady
state with power-law correlations
\begin{align*}
	G_+^<(\bar t \gg \bar x/u ;\bar x,0) = G_{0+}^<(0,\bar x)
\left\lvert {R_0}/{\bar x}\right\rvert^{{\tilde \gamma}^2}.	
\end{align*}
The corresponding momentum distribution function no longer exhibits a
discontinuity at $p=p_F$, but instead has a power-law singularity $\sim \lvert
p-p_F\rvert^{{\tilde\gamma}^2}$. This behavior is very similar to that observed
in an equilibrium Luttinger liquid~\cite{giamarchi-book},
however with an exponent ${\tilde \gamma}^2$ that differs from the equilibrium
one, $2\gamma = (1-K)^2/2K$. Hence, while interactions drive the dynamical
evolution which destroy the Fermi-liquid character of the spectral function, the
integrability prevents the system from relaxation into thermal equilibrium.

\section{Quench in the out-of-equilibrium Luttinger model}

In this section we consider the quench problem in the Luttinger liquid prepared
in a nonequilibrium initial state with double-step distribution functions.
First we present the key details of our calculations within the 
nonequilibrium bosonization framework and then discuss the obtained results.

\subsection{Solution via nonequilibrium bosonization}
\label{s3.1}

\begin{figure*}[t]
	\begin{center}
	\includegraphics[scale=0.175]{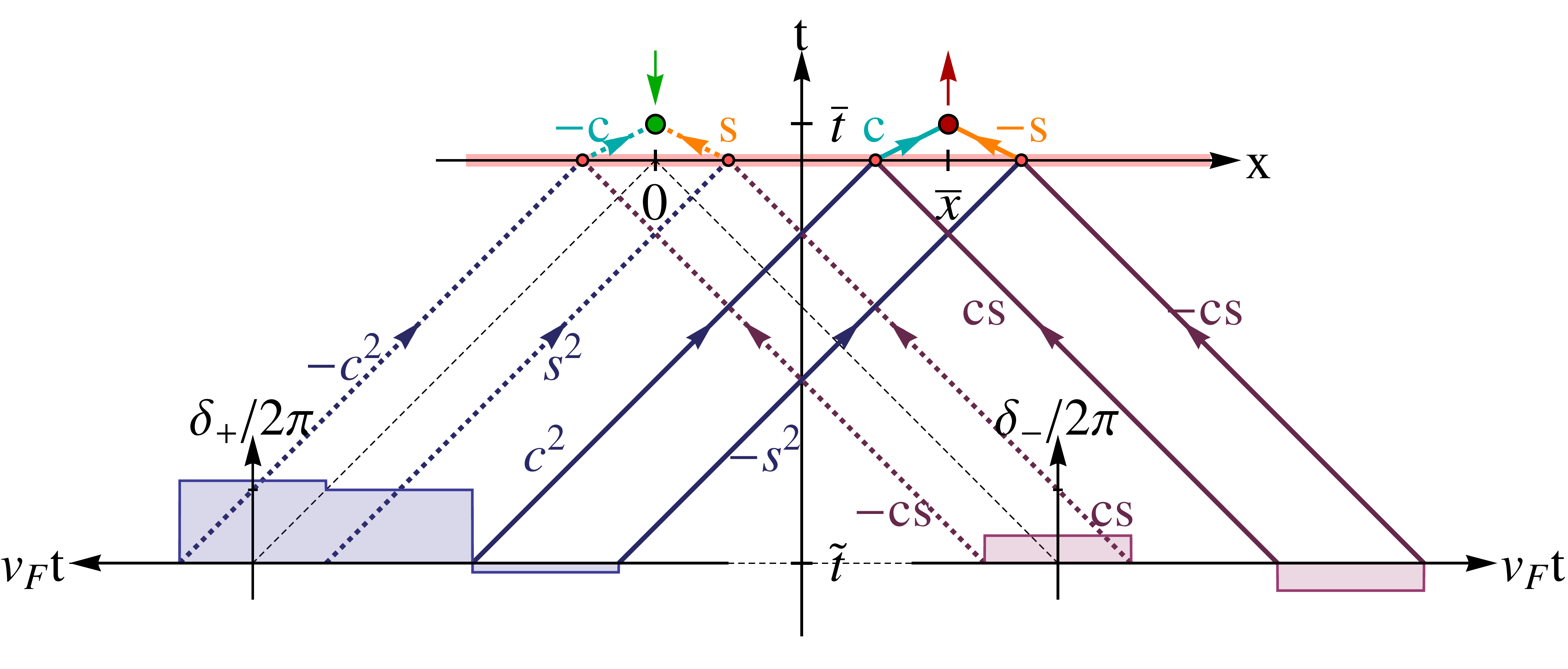}~\includegraphics[
scale=0.175]{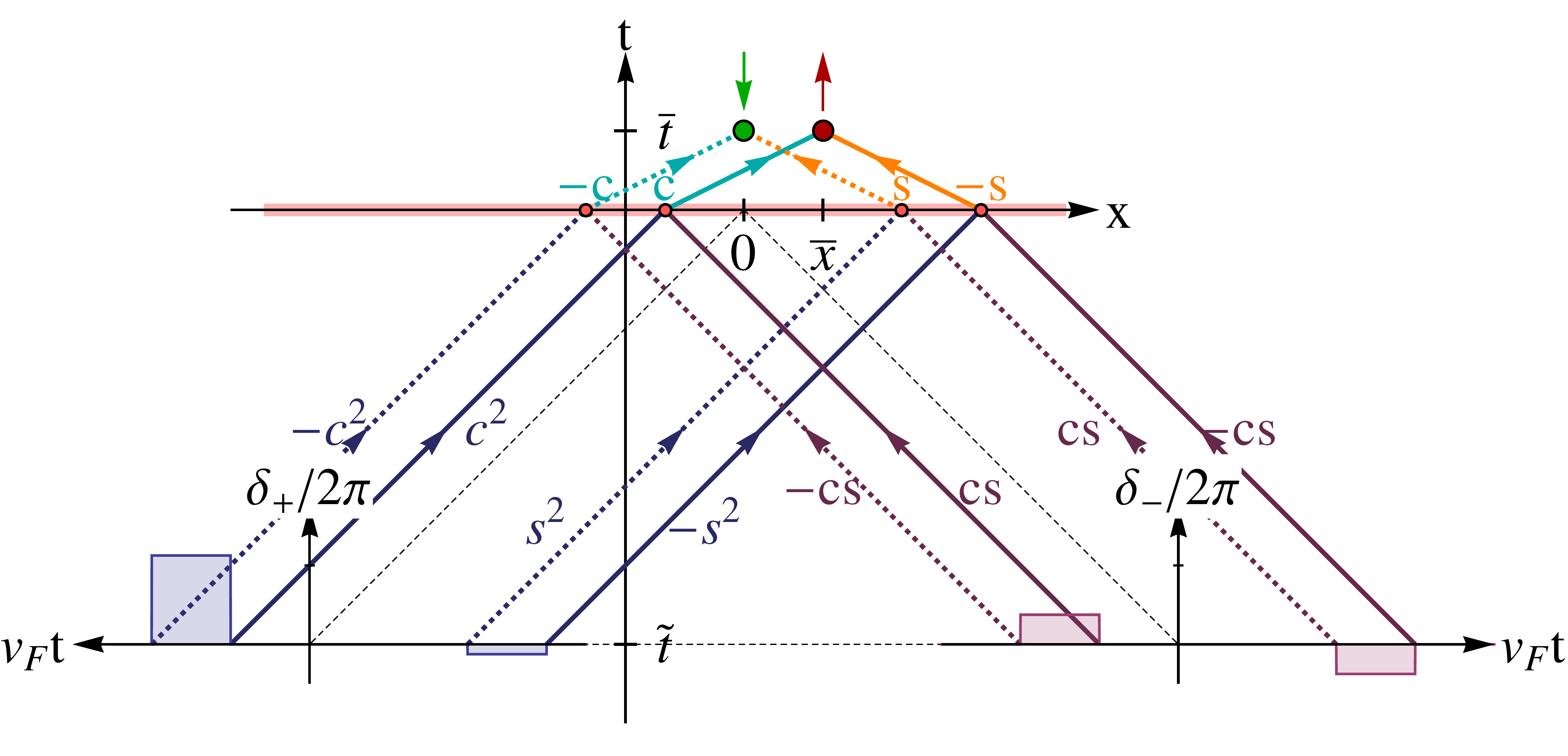}
	\end{center}
	\caption{Trajectories of density peaks and the corresponding counting
phases $\delta_\pm$ for the Green's function~(\ref{eq:GF_R}) in the case of
short times $2u \bar t<\bar x$ (\emph{left}) and long times $2u \bar t>\bar x>0$
(\emph{right}). The x-axis corresponds to the time $t=0$ when the quench takes
place. } \label{fig:DensityTrajPhases}
\end{figure*}

In our paper we consider initial states in which fermionic right-/left-moving
single-particle states $(\pm, \epsilon_i)$ are independently occupied according
to distribution functions
$f_\pm(\epsilon_i) = (1-a)\theta(-\epsilon_i) + a \theta( U
-\epsilon_i)$ which have two sharp Fermi edges at $\epsilon=0$ and $U$. 
For such nonequilibrium states 
the initial density matrix is not an exponential of an operator quadratic in the
plasmonic modes. This makes the nonequilibrium problem considerably more
complicated in comparison with the equilibrium one. 

A general framework to deal with this class of problems has been developed
in Ref.~\onlinecite{Gutman:2010} where the operator
bosonization method was combined with the Keldysh action technique. Within this
approach the right-movers' single-particle Green's function
\begin{align}
\label{eq:GF_R}
	iG^<_+(\bar x,\bar t;0,\bar t) &\equiv -\left\langle \psi_+^\dagger(0,\bar t) \psi_+(\bar x,\bar t)\right\rangle\\
	& = iG^<_{\eq +}(\bar x,\bar t;0,\bar t) \prod_{\eta=\pm}\Delta_\eta[\delta_\eta]/\Delta_0[\delta_\eta]\nonumber
\end{align}
can be expressed in terms of a functional determinant of a Fredholm operator,
\begin{align*}
	\Delta_\eta[\delta_\eta] = \Det\left[\xUnit-f_\eta+e^{i\delta_\eta} f_\eta\right].
\end{align*}
While the distribution function $f_\eta$ is diagonal in energy representation, the counting phase 
\begin{align}
	\delta_\eta(t) = 4\pi v_F \lim_{\tilde t\to -\infty} \int_{\tilde
t-t}^{0}\!\dd \tau\,\varrho_\eta^q(\eta v_F\tau,\tilde t)
\end{align}
is diagonal in the conjugate time representation. The equilibrium value $\Delta_0[\delta_\eta]$ of $\Delta_\eta[\delta_\eta]$ is obtained by replacing $f_\eta$ by the equilibrium distribution function $f_0(\epsilon)=\theta(-\epsilon)$. The counting phase is sensitive to the asymptotic behavior of the density trajectory $\varrho_\eta^q$, which is the advanced solution of the classical equations of motion
\begin{align}
	\left[\partial_t+\left(v_F+\frac{g_4}{2\pi}\right)\partial_x\right]
\varrho_+^q(x,t)+\frac{g_2}{2\pi} \partial_x \varrho_-^q(x,t)=& J(x,t),
\nonumber\\
	\left[\partial_t-\left(v_F+\frac{g_4}{2\pi}\right)\partial_x\right]
\varrho_-^q(x,t)-\frac{g_2}{2\pi} \partial_x \varrho_+^q(x,t)=& 0.
\end{align}
The source term $J(x,t) \equiv \frac 12 \delta(t-\bar t)
\left[\delta(x)-\delta(x-\bar x)\right]$ on the right-hand side of the equation
corresponds to the injection of a right-moving fermion at position $0$ and time
$\bar t$ and its removal at $(\bar x,\bar t)$. 
After the quench, $t>0$, the interaction couples right- and left-moving density
modes. To decouple them we apply the Bogoliubov
transformation~(\ref{eqn:Bogoliubov}), yielding
\begin{align}
	\left[\partial_t+u\partial_x\right] \tilde \varrho_+^q(x,t) =& c J(x,t),\nonumber \\
	\left[\partial_t-u\partial_x\right] \tilde \varrho_-^q(x,t) =& -s J(x,t). \label{eq:Charge_inj}
\end{align}
The charge configuration $\tilde \varrho^q_\eta$ describes
the ``advanced charge response''to the aforementioned injection and removal of
the right-moving fermion. 
It is a superposition of $\delta$-peaks moving
with velocities $\pm u$ after the quench and $\pm v_F$ prior to it. 

Solving the linear Eqs.~(\ref{eq:Charge_inj}) we find that at times $t>0$ after
the quench 
\begin{align*}
	\tilde \varrho^q_+(x,t) &= -\frac 12\ c\ \theta(\bar t-t) \\
&\times\left\lbrace \delta[x-u(t-\bar t)]-\delta[x-\bar x-u(t-\bar t)]\right\rbrace,\\
	\tilde \varrho^q_-(x,t) &= \frac 12\ s\ \theta(\bar t-t)\\
&\times \left\lbrace \delta[x+u(t-\bar t)]-\delta[x-\bar x+u(t-\bar t)]\right\rbrace.
\end{align*}
Requiring continuity at $t=0$ and inverting the Bogoliubov transformation we
obtain the charge density prior to the quench, $t<0$,
\begin{align*}
	\varrho^q_+(x,t) & = -\frac 12 \Bigl\lbrace c^2 \left(\delta[x+u \bar
t-v_F t]-\delta[x-\bar x+u \bar t-v_F t]\right) \\ & - s^2 \left(\delta[x-u \bar
t-v_F t]-\delta[x-\bar x-u \bar t-v_F t]\right)\Bigr\rbrace,\\
	\varrho^q_-(x,t) & = -\frac 12 cs \Bigl\lbrace\left(\delta[x+u \bar
t+v_F t]-\delta[x-\bar x+u \bar t+v_F t]\right) \\ &-  \left(\delta[x-u \bar
t+v_F t]-\delta[x-\bar x-u \bar t+v_F t]\right)\Bigr\rbrace.
\end{align*}
The latter yields the counting phases
\begin{align}
	\delta_+(t) &= 2\pi \Biggl\lbrace c^2 \left(\theta\left[t-\frac{-\bar
x+u\bar t}{v_F}\right]-\theta\left[t-\frac{u\bar t}{v_F}\right]\right) \\ &- s^2
\left(\theta\left[t-\frac{-\bar x-u\bar
t}{v_F}\right]-\theta\left[t-\frac{-u\bar t}{v_F}\right]\right)\Biggr\rbrace,
\nonumber\\
	\delta_-(t) &= -2\pi cs\ \Biggl\lbrace \left( \theta\left[t-\frac{-u\bar
t}{v_F}\right]-\theta\left[t-\frac{\bar x-u\bar
t}{v_F}\right]\right) \nonumber \\ & -\left(\theta\left[t-\frac{u\bar
t}{v_F}\right]-\theta\left[t-\frac{\bar x+u \bar
t}{v_F}\right]\right)\Biggr\rbrace.
\label{eq:Counting_phases}
\end{align}

We visualize the above solutions in Fig.~\ref{fig:DensityTrajPhases} which shows
the $\delta$-peak density trajectories in the $(x,t)$-plane. The filled circles
indicate the fermion injection and removal at time $\bar t$. Each fermion
fractionalizes into right- and left-moving plasmonic modes with weights $c$ and
$-s$. At the time of quench ($t=0$) the plasmonic peaks disintegrate into bare
particle-hole pair excitations (small circles on $x$-axis) which propagate with
smaller velocity $v_F$. The counting phases $\delta_\pm$ are calculated at
$\tilde t\to -\infty$.

The left panel of Fig.~\ref{fig:DensityTrajPhases} depicts the situation of
$2u\bar t< \lvert \bar x \rvert$ and the right panel corresponds to the case
$2u\bar t>\lvert \bar x\rvert$. In the second case the phases $\delta_\eta$
split into two independent pulses $\delta_\eta^{(1)}$, $\delta_\eta^{(2)}$ of
duration $\lvert \bar x\rvert /v_F$ which are shifted by $2u\bar t/v_F$. In the
long-time limit $2u\bar t\gg \lvert \bar x\rvert$, the coherence of right- and
left-moving plasmons is negligible and the determinant
$\Delta_\eta[\delta_\eta]\simeq
\Delta_\eta[\delta^{(1)}_\eta]\Delta_\eta[\delta_\eta^{(2)}]$ factorizes into
two single-pulse determinants which are of Toeplitz type.

Following Ref.~\onlinecite{Gutman:2011} we regularize the determinants
$\Delta_\eta[\delta_\eta]$ by introducing an ultraviolet cutoff $\Lambda$ and
discretize times in steps $\Delta t=\pi/\Lambda$. For counting phases
$\delta_\eta$ which are vanishing outside some time interval of length
$\tau$, the discretization gives rise to $N\times N$-matrices with
$N=\tau\Lambda/\pi$. Here we are interested in $\delta_\eta(t)$ which are
piecewise constant functions. Such dependence leads to the matrices of the
generalized Toeplitz form (see Appendix A). Various mathematical results exist
for the long-time asymptotic behavior of their determinants. In the simplest
situation that $\delta_\eta$ are single rectangular pulses [as e.g.\
$\delta_\eta^{(1)}(t)$] matrices are of Toeplitz form with symbols exhibiting
Fisher-Hartwig singularities. The Fisher-Hartwig conjecture~\cite{Its:2011} then
gives the leading exponential and power-law contribution to
$\Delta_\eta[\delta_\eta]$ for $N\gg 1$ (including numerical prefactors). The
extension proposed in Ref.~\onlinecite{Gutman:2011} allows for the calculation
of subleading power-law contributions. The phases $\delta_\pm(t)$ shown in
Fig.~\ref{fig:DensityTrajPhases} are not of a simple rectangular form. They
stem from the superposition of two rectangular pulses and therefore possess
four step-like discontinuities in time. This leads to a class of matrices
that are a generalization of Toeplitz matrices. The asymptotic behavior 
of the corresponding determinants can be found by a further generalization of
the Fisher-Hartwig conjecture (see Appendix A) which was put forward in
Ref.~\onlinecite{Protopopov:2012}. This conjecture was supported both 
by analytical~\cite{Protopopov:2012} and numerical~\cite{Protopopov:2012,
NgoDinh:2013} arguments. 

In Sec.~\ref{s3.2} we present and discuss the results for the Green's
function~(\ref{eq:GF_R}) obtained by means of the analysis of 
corresponding singular Fredholm determinants. The details
of these calculation can be found in Appendix B.

\subsection{Results}
\label{s3.2}

\begin{figure*}[t]
	\begin{center}
	\includegraphics[scale=0.4]{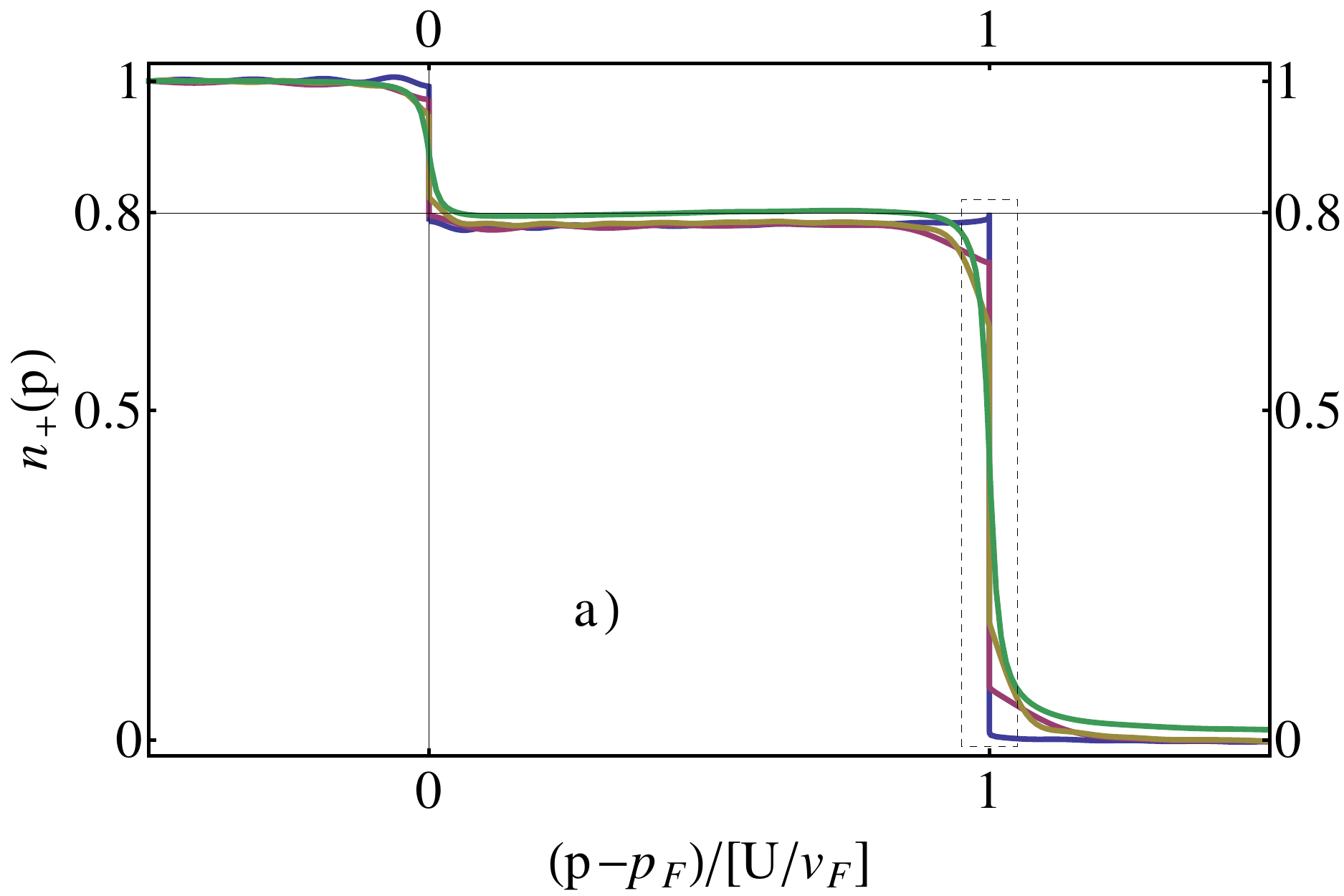}~\hspace{1.0cm}
\includegraphics[ scale=0.4]
{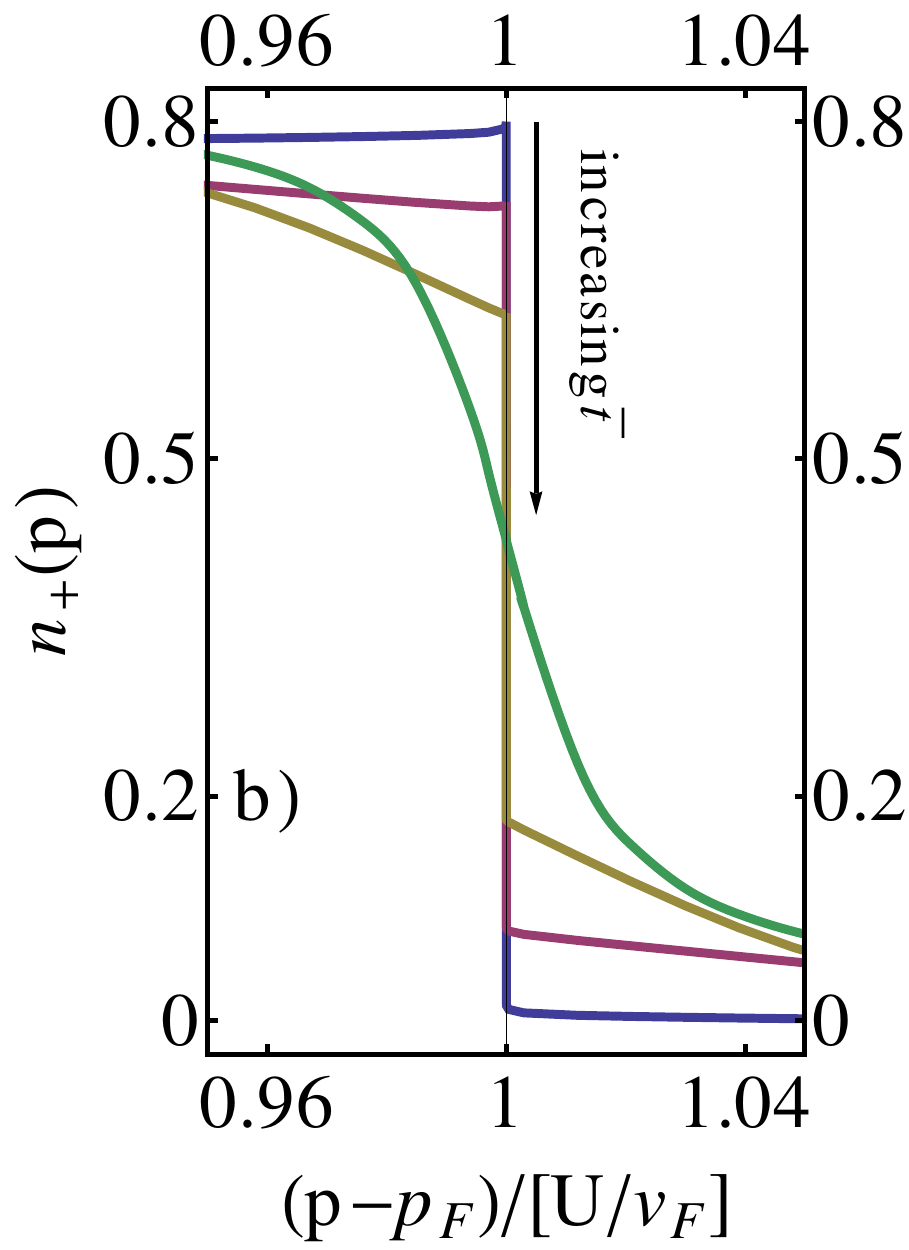}
	\end{center}
\label{fig:QuenchMomDistrKx8Ax8}
\caption{(a) The momentum distribution function for $K=0.8$ and
$a=0.8$ at times $ \bar t\, U  u/v_F =$1, 10, 25, and in the stationary
limit $\bar t \to \infty$, as obtained by a numerical evaluation of
the determinants in Eq.~(\ref{eq:GF_R}). Clear discontinuities at the edges
$p=p_F$ and $p=p_F+U/\vF$ [see zoom (b)] are visible which decrease with time
$\bar t$ according to Eq.~(\ref{eq:Z_12}) and eventually vanish.} 
\end{figure*}

In general the determinants entering Eq.~(\ref{eq:GF_R}) can be efficiently
evaluated numerically~\cite{Protopopov:2012,NgoDinh:2013}. For sufficiently
long times $u
\bar t \gg v_F/U$ analytical asymptotics can be found as we discuss below.

For an arbitrary relation between $\bar x$ and $u\bar t$ 
we have found that the Green's function is a linear combination of terms
involving different powers of $\lvert \bar x\rvert$, $\lvert 2 u\bar t+\bar
x\rvert$, $\lvert 2u\bar t-\bar x\rvert$, and $2u\bar t$. For simplicity we will
focus on two limiting cases: (i) long-distance behavior at finite times, $\bar x
\gg 2 u\bar t$,  and (ii) the stationary limit $\bar t\to \infty$.
We also assume a moderate repulsive interaction.

For finite times $\bar t$ and long distances $\bar x \gg u\bar t\gg \vF/U$
the Green's function is determined by two dominant terms,
\begin{widetext}
\begin{eqnarray}
G_+^<(\bar x,\bar t; 0,\bar t) =  G_{0+}^<(\bar x,0) e^{-\bar
t/(2\tau_\varphi)} \Biggl[\tilde\Gamma_1' \left(\frac{2u\bar
t\Lambda}{\pi\vF}\right)^{T_1'}\!\! \left(\frac{\pi U}{\Lambda}\right)^{V_1'}
+ \tilde\Gamma_2' e^{i\bar x U/\vF} \left(\frac{2u\bar
t\Lambda}{\pi\vF}\right)^{T_2'} \left(\frac{\pi
U}{\Lambda}\right)^{V_2'}\Biggr]
\end{eqnarray}
with exponents
\begin{align*}
	T'_j(V'_j) &= -\Re \Biggl[\left(\frac 32 -j
+s^2+2\beta\right)^2 +\left(s+2\tilde \beta\right)^2-\frac 14 \pm 2
c^2 s^2\Biggr], \quad j=1,2,
\end{align*}
\end{widetext}
and decay rate 
\begin{equation}
\label{eq:tau_phi}
 \tau_\varphi^{-1} = -4U \frac{u}{v_F} \Im(\beta+\tilde \beta),
\end{equation}
where
\begin{eqnarray}
\beta &=& \frac 1{2\pi i} \ln\left[a e^{-2\pi i
s^2}+1-a\right], \\
\tilde \beta &=& \frac 1{2\pi i} \ln\left[a e^{2\pi i cs}+1-a\right],
\end{eqnarray}
and $\tilde \Gamma'_j$ are numerical prefactors. 

Similar to the equilibrium quench, the entire $\bar x$-dependence of the
interacting Green's function after the quench is given by the noninteracting
factor, $G^>_{0+}\propto \bar x^{-1}$ so that correlations drop off with
distance in a Fermi-liquid-like manner. Correspondingly, the momentum
distribution function has discontinuities at $p=p_F$ and $p=p_F+U/v_F$,
signaling the existence of Landau quasiparticle states (see
Fig.~\ref{fig:QuenchMomDistrKx8Ax8} for $K=0.8$ and $a=0.8$). In the
nonequilibrium setup each of the two Fermi edges exhibits quasiparticles
with (in general different) weights 
\begin{equation}
\label{eq:Z_12}
Z_1\propto e^{-\bar t/(2\tau_\varphi)}\bar
t^{T'_1} U^{V'_1}, \quad Z_2\propto e^{-\bar t/(2\tau_\varphi)}\bar
t^{T'_2}
U^{V'_2}. 
\end{equation}
In striking contrast to the equilibrium situation, the quasiparticle
weights are not simply algebraically suppressed with time, but also exhibit
exponential decay with characteristic time $\tau_\varphi$ due to
nonequilibrium dephasing.

Let us now turn to the long-time limit, $\bar t\to \infty$. The quenched
system then relaxes to a stationary state without Fermi liquid
discontinuities, but with critical power-law correlations characteristic for
Luttinger liquid. In the case of moderate repulsive interaction
$\sqrt{2}-1\le K \le 1$ and at $\bar x\gg v_FU^{-1}$ leading 
contributions to the Green's function read
\begin{widetext}
\begin{align} \label{eqn:QuenchStatGF}
	G_+^<(\bar t \gg \bar x/ u;\bar x,0) = G_{0+}^<(0;\bar x)\ e^{-\kappa \lvert \bar x\rvert } \left(\tilde \Gamma_1 \left\lvert \frac{\Lambda\bar x}{\pi v_F}\right\rvert^{1+X_1} \left\lvert \frac{\pi U}{\Lambda}\right\rvert^{V_1} +\tilde \Gamma_2\ e^{iU\bar x/v_F} \left\lvert \frac{\Lambda\bar x}{\pi v_F}\right\rvert^{1+X_2} \left\lvert \frac{\pi U}{\Lambda}\right\rvert^{V_2}\right)
\end{align}
with exponents
\begin{align}
\label{eq:exp_X}
X_{j} (V_j) = -\frac12 {\rm Re}\left[\left(s^2-2\beta\right)^2-\left(s^2-2+j+2 \beta^\ast\right)^2
		- \left(c s-2 \tilde\beta\right)^2-\left(c
s-2\tilde\beta^\ast\right)^2\pm(c^2+s^2)^2\right], \quad j=1,2,
\end{align}
\end{widetext}
where star denotes the complex conjugation.
The decay length $\kappa^{-1}$ is equal to $\kappa^{-1} = 4u\tau_\varphi$ with
$\tau_\varphi$ given by Eq.~(\ref{eq:tau_phi}).
The numerical prefactors $\tilde \Gamma_j$ can be found in Appendix B, see
Eq.~(\ref{eq:Gamma_factors}).

\begin{figure}[t]
	\centering{\includegraphics[scale=0.225]{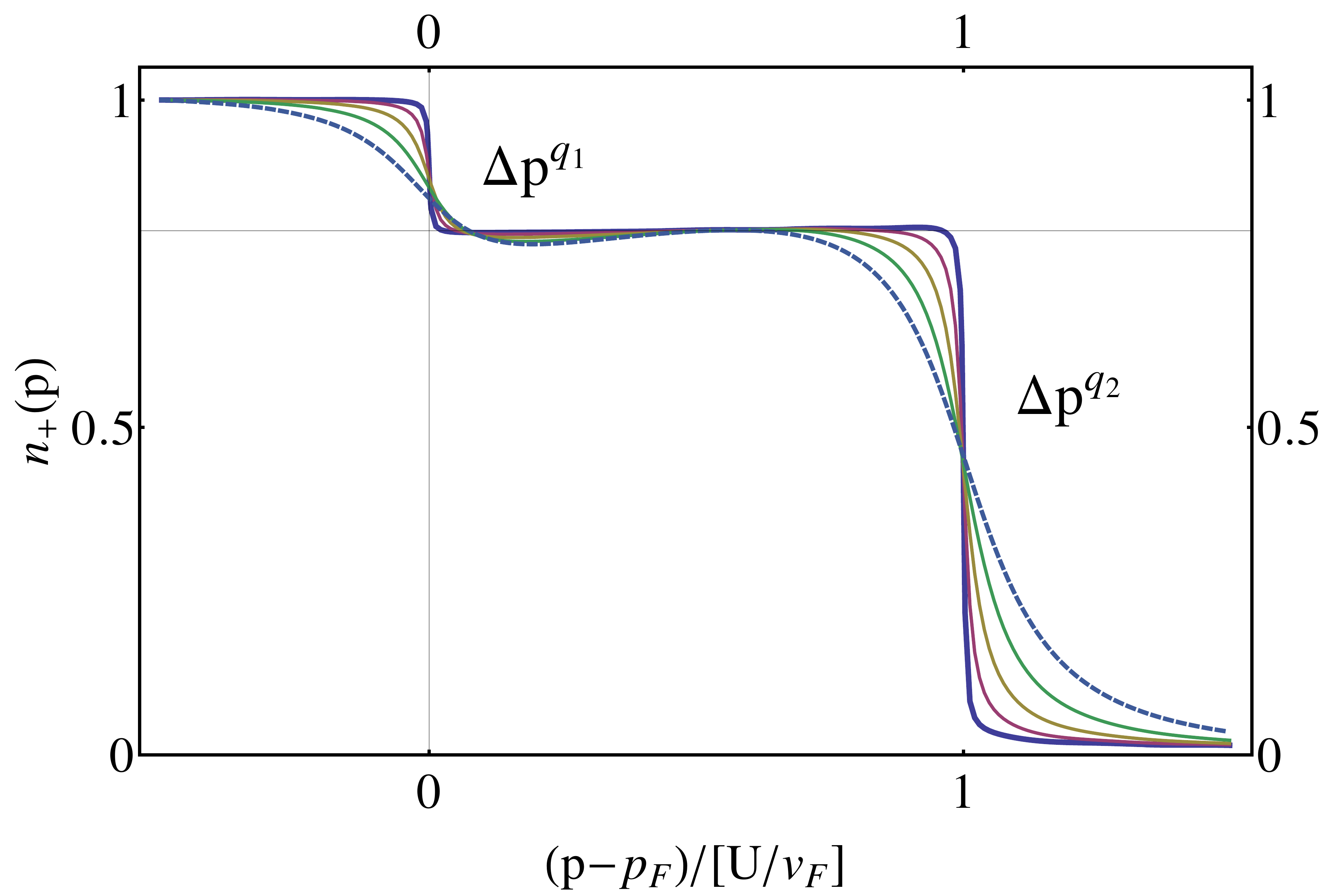}}

	\caption{Steady-state momentum distribution function for $a=0.8$ and
$K=$ 0.9 ({\it solid blue line}), 0.8, 0.7, 0.6, 0.5 ({\it dashed blue line})
obtained by a numerical evaluation of the determinants in Eq.~(\ref{eq:GF_R}).
At edges $p=p_F$ and $p=p_F+U/v_F$ the initial discontinuities
are replaced by power laws $\Delta p^{q_{1,2}}$ smeared by dephasing.}
\label{fig:quenchStatMomDistr}
\end{figure}

\begin{figure}[t]
	\centering{\includegraphics[scale=0.6]{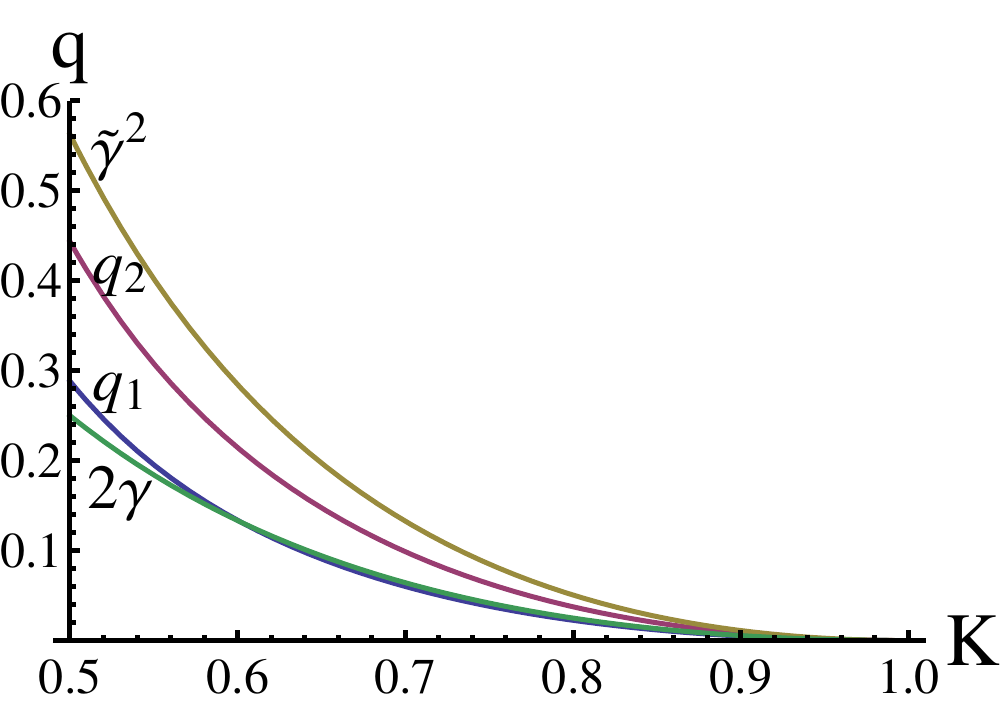}}
	\caption{Luttinger liquid exponents governing power-law singularities of
the momentum distribution functions: $q_1$ at $p=p_F$ and $q_2$ at $p=p_F+U/v_F$
for quenched nonequilibrium with $a=0.8$, $\tilde \gamma^2$ for quenched equilibrium, and
$2\gamma$ for equilibrium setup.} \label{fig:quenchStatExp}
\end{figure}

\begin{figure}[th]
	\centering{\includegraphics[scale=0.6]{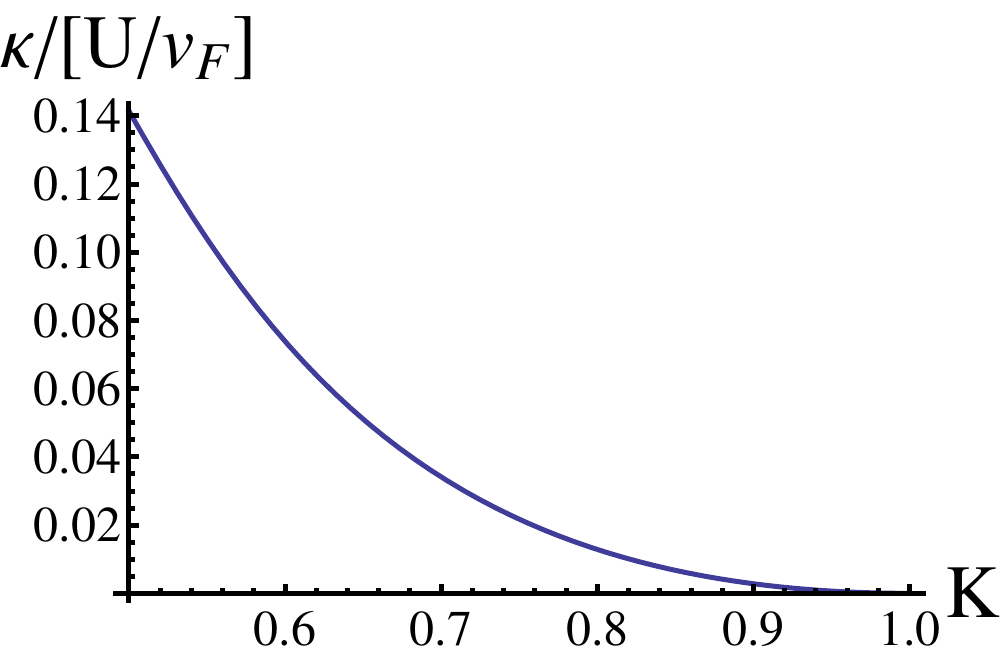}}
	\caption{Inverse decay length $\kappa$ in units of $U/v_F$ as a
function of the interaction constant $K$ shown for the double-step
distribution with $a=0.8$.} \label{fig:quenchStatDecayLength}
\end{figure}

Our results show that the limiting (long-time) stationary state retains
information about the system's prehistory, i.e. about the initial state prior to
the quench. Indeed, at long times after the quench the momentum distribution
function exhibits a double-step structure reminiscent of the original
nonequilibrium state 
(see Fig.~\ref{fig:quenchStatMomDistr}). The initial sharp discontinuities 
in $n(p)$ are replaced by power laws $\lvert p-p_F\rvert ^{q_1}$ and $\lvert
p-p_F-U/v_F\rvert ^{q_2}$ with $q_{1,2} = -(1 + X_{1,2})$ which are smeared by
nonequilibrium dephasing. The dependence of exponents $q_{1,2}$ and of the
inverse dephasing length $\kappa$ on the
interaction strength $K$ 
for a particular choice of the initial two-step distribution ($a=0.8$)
is shown in Figs.~\ref{fig:quenchStatExp} and \ref{fig:quenchStatDecayLength}. 

We stress that the exponents~(\ref{eq:exp_X}) differ from those in the
steady-state setup where the Luttinger liquid is coupled to noninteracting
reservoirs with nonequilibrium electron distributions~\cite{Gutman:2011}. 
Clearly, they also differ both from the equilibrium
exponent~\cite{giamarchi-book} and from that emerging after a quench in an
equilibrium Luttinger liquid, see Sec.~\ref{s2}.

\section{Summary}

To summarize, we have studied the dynamics of a nonequilibrium Luttinger liquid
after a sudden interaction quench by employing the nonequilibrium bosonization
formalism. At variance with a quench out of an initial equilibrium
state\cite{Cazalilla:2006,Iucci:2009}, the
quasiparticle weights decay exponentially with time after the quench. This
exponential behavior is however not a relaxation into a thermal state, which is
related to the integrability of the model.
In particular, for an initial distribution with two Fermi edges, the
distribution emerging at long times retains a double-step structure, with
power-law singularities smeared by dephasing. The corresponding exponents as
well as the dephasing rate depend on the initial nonequilibrium state. 

In conclusion we mention that the framework developed here is
also applicable to a more general situation when the interaction
region possesses both spatial and temporal boundaries. In this
case the counting phases will be determined by the 
fractionalization processes on all the boundaries.     

\section{Acknowledgement}

A.D.M. thanks M.A.~Cazalilla for an instructive discussion.
This work was supported by the collaborative research grant SFB/TR12 of the DFG
and by German-Israeli Foundation. D.B. is grateful to the TKM institute at KIT
for its hospitality.  

\appendix

\section{Generalized Toeplitz determinants}
In this appendix we summarize the main results of Ref.~\onlinecite{Protopopov:2012}
for the asymptotic behavior of generalized Toeplitz determinants. 
We consider a generalized Toeplitz matrix
\begin{equation}
\label{eq:def_gen_TM}
g_{j,k}=\int_{-\Lambda}^\Lambda\!\frac{d \epsilon}{2\Lambda}\, e^{-i\epsilon \pi/\Lambda [j-k-\delta(t_j)/(2\pi)]}
\tilde g(t_j,\epsilon)\end{equation}
which is defined via its symbol
\begin{equation}
\label{eq:symbol_gen_TM}
\tilde g(t,\epsilon)\equiv1+\left(e^{i\delta(t)}-1\right)f(\epsilon).
\end{equation}
Let us focus on the special case when both
the phase $\delta(t)$ and the distribution function $f(\epsilon)$ are
piecewise constant functions
with jumps at times $\tau_1<\tau_2<\ldots<\tau_{N_\tau}$ and energies
$\mu_1<\mu_2<\ldots <\mu_{N_\mu}$, 
respectively. They satisfy the boundary conditions $\delta(t)=0$ for
$t\notin[\tau_1,\tau_{N_\tau}]$, $f(\epsilon)=1$ for $\epsilon<\mu_1$, and 
$f(\epsilon)=0$ for $\epsilon>\mu_{N_\mu}$.
The discontinuity points define a grid which subdivides the time-energy plane
in domains with different values of the symbol. The domains can be labeled by   
the time indices $j\in \lbrace 0,\ldots, N_\tau\rbrace$, 
and energy indices $k\in\lbrace 0,\ldots, N_\mu\rbrace$.
One associates with this set of domains a set of number~$c_{jk}$, 
\begin{eqnarray} \label{eqn:defCjk}
	c_{jk} &=& \frac 1{2\pi i} \ln \tilde g(\tau_j+0,\mu_k+0)+n_{jk},\\
	c_{j0} &=& \delta(t_j+0)/(2\pi),\quad
c_{0k}=c_{N_\tau,k}=c_{j,N_\mu}=0. \nonumber
\end{eqnarray}
where $\lbrace n_{jk}\rbrace$ is an arbitrary set of integers. 
In the above equation the logarithm $\ln\tilde g$ is
understood as evaluated at its principal branch, $\Im\ln \tilde g \in
(-\pi,\pi]$. The summation over integers $n_{jk}$ 
hence amounts to summing over different branches of the logarithms.

It was conjectured in Ref.~\onlinecite{Protopopov:2012} that
the asymptotic behavior of the (normalized) determinant 
$\bar\Delta[\delta(t),f(\epsilon)] = \Delta[\delta(t),f(\epsilon)]/\Delta[\delta(t),T=0]$
takes the form
\begin{widetext}
\begin{equation}
\label{eq:non_T_norm_series}
	\bar\Delta[\delta(t),f(\epsilon)] =\sum_{\lbrace n_{jk}\rbrace} \bar\Gamma_{\lbrace n_{jk}\rbrace}\ 
\exp\Biggl[i\sum_{1 \leq j \leq N_t} \sum_{1\leq k \leq N_\mu} \tau_j \beta_{jk} \mu_k  \Biggr]
		\prod_{1\le j<l\leq N_t}\prod_{1\le k<m\leq N_\mu}\ \Bigl[(\tau_l-\tau_j)(\mu_k-\mu_m)\Bigr]^{\gamma_{jl,km}},
\end{equation}
where
\begin{equation}
\gamma_{jl,km}= -c_{jk} c_{lm} - c_{jm} c_{lk}.
\end{equation}
The normalized determinant is cutoff ($\Lambda$) independent. All dependence on $\Lambda$ comes from
the zero temperature determinant,  which up to a constant prefactor reads
\begin{equation}
\label{eq:Det_T0}
\Delta[\delta,T=0]=\exp\Biggl[-i\sum_{1 \leq j \leq N_t} \Lambda\tau_j\frac{(\delta_j-\delta_{j-1})}{2\pi}\Biggr] 
\prod_{1\le j<l\leq N_t}\left\lvert \frac{\Lambda\left(\tau_j-\tau_l\right)}{\pi}\right\rvert^{
{(\delta_j-\delta_{j-1})(\delta_l-\delta_{l-1})}/{4\pi^2}}, \nonumber
\end{equation}
where we have defined the phases $\delta_j \equiv \delta(t_j+0)$. 
While a rigorous mathematical proof of these results is still missing, there exists a strong 
analytical~\cite{Protopopov:2012}  and numerical~\cite{Protopopov:2012, NgoDinh:2013} evidence in their favour.
The above asymptotic expressions for the generalized Toeplitz
determinants are used in Appendix B for evaluation of the Green's function 
in the nonequilibrium Luttinger liquid after the quench.

\section{Asymptotic behavior of the Green's function}

This appendix contains details of calculations of the asymptotic behavior of
the determinants that lead to the results presented in Sec.\ref{s3.2}. 

Throughout the paper we use the nonequilibrium version of canonical bosonization
developed
in Refs.~\onlinecite{Gutman:2010,Gutman:2011,Protopopov:2012,Protopopov:2013}.
In this framework nonequilibrium corrections to the right-movers' equal-time Green's function $G_{+}^\gtrless(\bar t;\bar x,0)$ [see Eq.(\ref{eq:GF_R})] are expressed in terms of Fredholm determinants
\begin{align*}
	\Delta_\mu\equiv \Det\left[\xUnit+\left(e^{i\delta_\mu}-\xUnit\right)f_\mu\right].
\end{align*}
with the counting phases $\delta_\mu$ [Eq.(\ref{eq:Counting_phases})] which were found in section~\ref{s3.1}.

Let us consider first the equilibrium situation with distribution functions 
$f_+(\epsilon)=f_-(\epsilon)=f_0(\epsilon)=\theta(-\epsilon)$.
Equations~(\ref{eq:non_T_norm_series}) and (\ref{eq:Det_T0}) yield
\begin{align*}
	\Delta_+ &= G(1-s^2)G(1+s^2)G(1-c^2) G(1+c^2)\ e^{i\Lambda \bar x/v_F} \left\lvert\frac{\Lambda \bar x}{\pi v_F}\right\rvert^{-c^4-s^4} \left\lvert \frac{(2u\bar t)^2-\bar x^2}{(2u\bar t)^2}\right\rvert^{s^2c^2},\\
	\Delta_- &= G(1-cs)^2G(1+cs)^2\ \left\lvert\frac{\Lambda \bar x}{\pi v_F}\right\rvert^{-2c^2s^2} \left\lvert \frac{(2u\bar t)^2-\bar x^2}{(2u\bar t)^2}\right\rvert^{s^2c^2}.
\end{align*} 
The prefactors containing the Barnes G-functions $G$ do not directly follow
from the asymptotic formulae for the generalized Toeplitz determinants.
They can be found from the long-time limit where the factorization into simple
Toeplitz determinants is applicable (we refer the reader to 
Ref.~\onlinecite{Gutman:2011} for more details).

The Green's function following the equilibrium interaction quench is thus
\begin{align*}
	G^<_+(\bar t;\bar x,0) \propto e^{i\Lambda \bar x/v_F} \left\lvert\frac{\Lambda \bar x}{\pi v_F}\right\rvert^{-(c^2+s^2)^2} \left\lvert \frac{(2u\bar t)^2-\bar x^2}{(2u\bar t)^2}\right\rvert^{2 s^2c^2}.
\end{align*}
Since $2 c s=\tilde \gamma$ and $-(c^2+s^2)^2=-1-\tilde\gamma^2$ the power laws are in perfect agreement with the exact result (\ref{eqn:LLEqQGF}) of Ref.~\onlinecite{Cazalilla:2006}.
In the following we will use the equilibrium quench as reference case to normalize our Fredholm determinants.

The situation becomes more complicated when turning to nonequilibrium. We 
consider the double step distribution functions $f_\pm(\epsilon) =
(1-a)\theta(-\epsilon) + a \theta( U
-\epsilon)$ for right- and left-movers as
the initial steady state of the noninteracting Fermi sea before the quench.
Using the asymptotic formulae (\ref{eq:non_T_norm_series}) and
(\ref{eq:Det_T0})  we obtain
\begin{align*}
	G_+^<(\bar t;\bar x,0) = G_{0+}^<(0,\bar x)\ \tilde \Delta_+\ \tilde \Delta_-
\end{align*}
for $\lvert \bar x\rvert, \lvert 2u\bar t-\lvert \bar x\rvert\rvert, \lvert 2u\bar t\rvert \gg v_F/U$ with equilibrium-normalized determinants
\begin{align*}
	\tilde \Delta_+ = &\sum_{n_1,n_2,n_3 \in \mathbb{Z}}\,\tilde\Gamma_+(n_1,n_2,n_3)\, e^{i(\beta_1+\beta_2)U \lvert \bar x\rvert /v_F} e^{i(n_1+n_2-n_3)U \lvert \bar x\rvert /v_F} e^{i n_3 2U \bar t\, u/v_F}\\
	 & \times \left\lvert\frac{\Lambda}{\pi v_F}\right\rvert^{1+\gamma_++V_+} \lvert \bar x\rvert^{1+X_{+}}\ \left(2u\bar t-\lvert \bar x\rvert\right)^{D_+}\left(2u\bar t+\lvert \bar x\rvert \right)^{S_+} (2u\bar t)^{T_+} \left\lvert \frac{\pi U}\Lambda\right\rvert^{V_+},\\
	 \tilde \Delta_- =& \sum_{n_4,n_5,n_6 \in \mathbb{Z}}\,\tilde\Gamma_-(n_4,n_5,n_6)\, e^{i(\beta_3+\beta_4)U\lvert \bar x\rvert /v_F} e^{i(n_4+n_5-n_6)U \lvert \bar x\rvert /v_F} e^{i n_6 2U \bar t\, u/v_F}\\
	 & \times \left\lvert\frac{\Lambda}{\pi v_F}\right\rvert^{\gamma_-+V_-} \lvert \bar x\rvert^{X_{-}}\ \left(2u\bar t-\lvert \bar x\rvert\right)^{D_-}\left(2u\bar t+\lvert \bar x\rvert \right)^{S_-} (2u\bar t)^{T_-} \left\lvert \frac{\pi U}\Lambda\right\rvert^{V_-}
\end{align*}
for $2u\bar t>\bar x>0$ and 
\begin{align*}
	\tilde \Delta_+ = &\sum_{n_1,n_2,n_3 \in \mathbb{Z} }\,\tilde\Gamma'_+(n_1,n_2,n_3)\, e^{i(\beta_1+\beta_2) 2U\bar t u/v_F} e^{i(n_1+n_2-n_3)U \lvert \bar x\rvert /v_F} e^{i n_3 2U \bar t\, u/v_F}\\
	 & \times \left\lvert\frac{\Lambda}{\pi v_F}\right\rvert^{1+\gamma'_++V'_+} \lvert \bar x\rvert^{1+X'_{+}}\ \left(-2u\bar t+\lvert \bar x\rvert\right)^{D'_+}\left(2u\bar t+\lvert \bar x\rvert \right)^{S'_+} (2u\bar t)^{T'_+} \left\lvert \frac{\pi U}\Lambda\right\rvert^{V'_+},\\
	 \tilde\Delta_- =& \sum_{n_4,n_5,n_6 \in \mathbb{Z}}\,\tilde\Gamma'_-(n_4,n_5,n_6)\, e^{i(\beta_3+\beta_4) 2U\bar t u/v_F} e^{i(n_4+n_5-n_6)U \lvert \bar x\rvert/v_F} e^{i n_6 2U \bar t\, u/v_F}\\
	 & \times \left\lvert\frac{\Lambda}{\pi v_F}\right\rvert^{\gamma'_-+V'_-} \lvert \bar x\rvert^{X'_{-}}\ \left(-2u\bar t+\lvert \bar x\rvert\right)^{D'_-}\left(2u\bar t+\lvert \bar x\rvert \right)^{S'_-} (2u\bar t)^{T'_-} \left\lvert \frac{\pi U}\Lambda\right\rvert^{V'_-}
\end{align*}
for $0<2u\bar t < \bar x$. Here, we left the $n_j$-dependence of the exponents
$X_\pm, T_\pm,\ldots$ implicit. $\tilde \Gamma_\pm, \tilde \Gamma'_\pm$ are 
numerical prefactors which are not known in general. The determinants for $\bar
x<0$ are obtained by complex conjugation.

The exponents differ in the two regimes $2u\bar t\gtrless \lvert \bar x\rvert$, which we consider below
separately.

\subsection*{Regime of separated phase pulses, $2u\bar t>\lvert \bar  x\rvert$}
Here the exponents are
\begin{align*}
	X_+ =& \left(-\beta _2+c^2-n_2\right) \left(\beta _2-c^2+n_2-n_3\right)+\left(-\beta _1-n_1+n_3\right) \left(\beta _1+n_1\right)\\
	&+\left(-\beta _2-n_2+n_3\right) \left(\beta _2+n_2\right)+\left(-\beta _1-n_1-s^2\right) \left(\beta _1+n_1-n_3+s^2\right),\\
	T_+ = &\left(-n_1+n_3-s^2-\beta_1\right) \left(c^2-n_2-\beta_2\right)+\left(-n_1-s^2-\beta_1\right) \left(c^2-n_2+n_3-\beta_2\right)
	\\ & +\left(n_1-n_3+\beta_1\right) \left(n_2+\beta_2\right) +\left(n_1+\beta_1\right) \left(n_2-n_3+\beta_2\right),\\
	D_+= &\left(n_1-n_3+\beta_1\right) \left(-n_2+n_3-\beta_2\right)+\left(-n_1+n_3-s^2-\beta_1\right) \left(-c^2+n_2-n_3+\beta_2\right),\\
	S_+ = &\left(\beta _1+n_1\right) \left(c^2-2 \left(\beta _2+n_2\right)\right)-s^2 \left(\beta _2-c^2+n_2\right),\\
	\gamma_+ = &-c^4-s^4,\\
	X_- = &\left(-n_4+n_6-\beta_3\right) \left(n_4+\beta_3\right)+\left(-n_4+c s-\beta_3\right) \left(n_4-n_6-c s+\beta_3\right)\\
			&+\left(-n_5+n_6-\beta_4\right) \left(n_5+\beta_4\right)+\left(-n_5-c s-\beta_4\right) \left(n_5-n_6+c s+\beta_4\right),\\
	T_- = &\left(-n_4+n_6+c s-\beta_3\right) \left(-n_5-c s-\beta_4\right)+\left(-n_4+c s-\beta_3\right) \left(-n_5+n_6-c s-\beta_4\right)\\
		&+\left(n_4-n_6+\beta_3\right) \left(n_5+\beta_4\right)+\left(n_4+\beta_3\right) \left(n_5-n_6+\beta_4\right),\\
	D_- = &\left(n_4-n_6+\beta_3\right) \left(-n_5+n_6-\beta_4\right)+\left(-n_4+n_6+c s-\beta_3\right) \left(n_5-n_6+c s+\beta_4\right),\\
	S_- = &c s \left(n_5+c s+\beta_4\right)+\left(n_4+\beta_3\right) \left(-c s-2 \left(n_5+\beta_4\right)\right),\\
	\gamma_- =& -2 c^2 s^2.
\end{align*}
In the long-time limit $2u\bar t\gg \lvert \bar x\rvert$, the powers simplify to $\left(2u\bar t-\lvert \bar x\rvert\right)^{D_\pm}\left(2u\bar t+\lvert \bar x\rvert \right)^{S_\pm} (2u\bar t)^{T_\pm} \to (2u\bar t)^{\tilde T_{\pm}}$ with $\tilde T_+=-2n_3^2\le 0$ and $\tilde T_-=-2 n_6^2\le 0$. Thus the correlation function relaxes to a stationary solution where solely terms with $n_3=0=n_6$ contribute. The remaining powers simplify to
\begin{align*}
	X_+ =& -2 \left(n_1-\frac{-s^2-2\beta_1}2\right)^2-2 \left(n_2-\frac{c^2-2 \beta_2}2\right)^2-\frac{c^4+s^4}2,\\
	V_+ =& -2 \left(n_1-\frac{-s^2-2\beta_1}2\right)^2-2 \left(n_2-\frac{c^2-2 \beta_2}2\right)^2+\frac{c^4+s^4}2,\\
	X_- =& -2 \left(n_4-\frac{c s-2 \beta_3}2\right)^2-2 \left(n_5-\frac{-c s-2\beta_4}2\right)^2- c^2s^2,\\
	V_- =& -2 \left(n_4-\frac{c s-2 \beta_3}2\right)^2-2 \left(n_5-\frac{-c s-2\beta_4}2\right)^2+ c^2s^2.
\end{align*}
Since in the long-time limit the phases split into independent pulses, all
Fredholm determinants factorize into Toeplitz determinants and the prefactors
can be found in the closed analytical form using the generalized
Fisher-Hartwig formula~\cite{Gutman:2011}
\begin{align}
	\tilde \Gamma_+(n_1,n_2,n_3=0) = &
\frac{G(1-s^2-\beta_1-n_1)G(1+s^2+\beta_1+n_1)G(1+\beta_1+n_1)G(1-\beta_1-n_1)}{
G(1-s^2) G(1+s^2)}\nonumber\\
	&\times 
\frac{G(1+c^2-\beta_2-n_2)G(1-c^2+\beta_2+n_2)G(1+\beta_2+n_2)G(1-\beta_2-n_2)}{
G(1+c^2) G(1-c^2)},\nonumber\\
	\tilde \Gamma_-(n_4,n_5,n_6=0) = &
	\frac{
G(1+cs-\beta_3-n_4)G(1-cs+\beta_3+n_4)G(1+\beta_3+n_4)G(1-\beta_3-n_4)}{
G(1+cs)G(1-cs)}\nonumber\\
		& \times
\frac{G(1-cs-\beta_4-n_5)G(1+cs+\beta_4+n_5)G(1+\beta_4+n_5)G(1-\beta_4-n_5)}{
G(1-cs)G(1+cs)}.
\label{eq:Gamma_factors}
\end{align}
For \emph{moderate} repulsive interaction $\sqrt{2}-1\le K \le 1$, the dominant
powers $\Re X_\pm$ are due to $(n_1,n_2)=(0,1),(0,0)$ and $(n_4,n_5)=(0,0)$.
These contributions are taken into account in (\ref{eqn:QuenchStatGF}) with
$\tilde \Gamma_1\equiv \tilde \Gamma_+(0,0,0)\tilde \Gamma_-(0,0,0)$, $\tilde
\Gamma_2\equiv \tilde \Gamma_+(0,1,0)\tilde \Gamma_-(0,0,0)$.
In the equilibrium limit, $a\to 0$, prefactors vanish for all $n_j$ but
$n_1=n_2=n_4=n_5=0$ for which one recovers the equilibrium exponents.

\subsection*{Regime of overlapping phase pulses, $2u\bar t<\lvert \bar 
x\rvert$}
Here the exponents are
\begin{align*}
	X'_+ =& \left(-c^2+n_2-n_3-\beta_1\right) \left(c^2-n_2-\beta_2\right)+\left(-n_1-s^2-\beta_1\right) \left(n_1-n_3+s^2-\beta_2\right)
	 \\ &+\left(-n_2+n_3+\beta_1\right) \left(n_2+\beta_2\right)+\left(n_1+\beta_1\right) \left(-n_1+n_3+\beta_2\right),\\
	T'_+  = &\left(n_2-n_3-\beta_1\right) \left(n_1+\beta_1\right)+\left(-n_1-s^2-\beta_1\right) \left(c^2-n_2+n_3+\beta_1\right)
	\\ & +\left(n_1-n_3-\beta_2\right) \left(n_2+\beta_2\right)+\left(c^2-n_2-\beta_2\right) \left(-n_1+n_3-s^2+\beta_2\right),\\
	D'_+ = &\left(-n_2+n_3+\beta_1\right) \left(n_1-n_3-\beta_2\right)+\left(-c^2+n_2-n_3-\beta_1\right) \left(-n_1+n_3-s^2+\beta_2\right),\\
	S'_+ = &-s^2 \left(-c^2+n_2+\beta_2\right)+\left(n_1+\beta_1\right) \left(c^2-2 \left(n_2+\beta_2\right)\right),\\
	X'_- = &\left(n_5-n_6-c s-\beta_3\right) \left(-n_5+c s-\beta_4\right)+\left(-n_4-c s-\beta_3\right) \left(n_4-n_6+c s-\beta_4\right)
	\\& +\left(-n_5+n_6+\beta_3\right) \left(n_5+\beta_4\right)+\left(n_4+\beta_3\right) \left(-n_4+n_6+\beta_4\right),\\
	T'_- = &\left(n_5-n_6-\beta_3\right) \left(n_4+\beta_3\right)+\left(-n_4-c s-\beta_3\right) \left(-n_5+n_6+c s+\beta_3\right)
	\\ &+\left(n_4-n_6-\beta_4\right) \left(n_5+\beta_4\right)+\left(-n_5+c s-\beta_4\right) \left(-n_4+n_6-c s+\beta_4\right),\\
	D'_- = &\left(-n_5+n_6+\beta_3\right) \left(n_4-n_6-\beta_4\right)+\left(n_5-n_6-c s-\beta_3\right) \left(-n_4+n_6-c s+\beta_4\right),\\
	S'_- = &-c s \left(n_5-c s+\beta_4\right)+\left(n_4+\beta_3\right)
\left(c s-2 \left(n_5+\beta_4\right)\right).
\end{align*}

For long distances $\lvert \bar x \rvert \gg 2u\bar t$ the power-law dependence on distance simplifies to\\ $\lvert \bar x\rvert^{X'_{\pm}}\ \left(-2u\bar t+\lvert \bar x\rvert\right)^{D'_\pm}\left(2u\bar t+\lvert \bar x\rvert \right)^{S'_\pm} \to \lvert \bar x \rvert ^{\tilde X_\pm}$ with the exponents 
\begin{align*}
	\tilde X_+ = -2(n_3+1/2-n_1-n_2)^2 - \frac 12,\quad \tilde X_- = -2(n_6-n_4-n_5)^2.
\end{align*}
For $\lvert \bar x\rvert \to \infty$ all terms vanish except for $n_3=n_1+n_2$ or $n_3=n_1+n_2-1$, and $n_6=n_4+n_5$. Then $1+\tilde X_+=0=\tilde X_-$, i.e.\ the normalized determinants $\tilde \Delta_\pm$ are independent of $\bar x$, and correlations drop off like $G_+^<(\bar t;\bar x,0)\sim G_{0+}^<(\bar t;\bar x,0)\sim \bar x^{-1}$.

The remaining exponents are
\begin{align}
\begin{split} \label{eqn:quenchLongXExpR0}
	T'_+ (V'_+) &= -2 \left(n_1-\frac{-1/2-s^2-2\beta_1}2\right)^2-2 \left(n_2-\frac{-1/2+c^2-2\beta_2}2\right)^2+\frac 14\mp c^2 s^2
\end{split}
\end{align}
for $n_3=n_1+n_2$,
\begin{align}
\begin{split} \label{eqn:quenchLongXExpR1}
	T'_+(V'_+) &= -2 \left(n_1-\frac{1/2-s^2-2\beta_1}2\right)^2-2 \left(n_2-\frac{1/2+c^2-2\beta_2}2\right)^2+\frac 14\mp c^2 s^2
	\end{split}
\end{align}
for $n_3=n_1+n_2-1$
and 
\begin{align}
\begin{split} \label{eqn:quenchLongXExpL}
	T'_- (V'_-) &=-2\left(n_4-\frac{-cs-2\beta_3}2\right)^2-2\left(n_5-\frac{cs-2\beta_4}2\right)^2\mp c^2s^2
\end{split}
\end{align}
for $n_6=n_4+n_5$.
\end{widetext}

\end{document}